\documentclass[11pt,a4paper,numcompress]{article}
\usepackage{jinstpub}
\usepackage{graphicx}
\usepackage{subcaption}  

\usepackage{subcaption}
\usepackage{lineno}
\usepackage{setspace}
\usepackage[dvipsnames]{xcolor}

\title{Machine Learning-Based Cluster Classification to Suppress Background in a Prototype RPC Detector}

\author[a,b,1]{Souvik Chattopadhay,\note{Corresponding author.}}

\author[a,b]{Zubayer Ahammed}

\affiliation[a]{Variable Energy Cyclotron Centre,
 1/AF,Bidhannagar, Kolkata, India}
\affiliation[b]{Homi Bhabha National Institute,
BARC Training School Complex, Mumbai, India}
\emailAdd{souvikcjee@gmail.com}

\abstract{Resistive Plate Chambers (RPCs) are widely used as tracking detectors in many high-energy physics experiments. It has been observed that low-resistive bakelite RPC prototypes frequently exhibit a secondary hit component, appearing as a long tail or an additional peak in the time-correlation spectra relative to the trigger detector. These secondary hits, which affect both the time and spatial resolution, are difficult to distinguish from genuine signals in high-rate environments without an external trigger. As a result, they can significantly degrade track reconstruction efficiency and increase processing time. We present a machine-learning–based strategy to separate signal and background hit clusters using fifteen cluster-level descriptors that encode both statistical properties (histogram mean, width, cluster size) and fit-based parameters (Gaussian-fit mean, width, amplitude, $\chi^2$, NDF) of the time and ADC distributions. Using laboratory data collected from a single-gap low resistive RPC with a three-scintillator master trigger, we trained and evaluated three classifiers-DNN, 1D-CNN, and XGBoost-on balanced signal/background samples. All models demonstrate strong discrimination capability, with XGBoost showing the most robust generalization performance. Feature-importance analysis indicates that cluster size and temporal-shape descriptors are the dominant discriminants. These results highlight that compact, interpretable cluster-level features combined with machine-learning classifiers offer a practical and effective approach to suppress background in self-triggering low resistive RPC detectors.}
\begin{document}
\maketitle

\section{Introduction}
\label{sec:intro}
Resistive Plate Chambers (RPCs) \cite{RPCSANTONICO} have been widely used as timing and triggering detectors in high-energy physics experiments for more than two decades \cite{atlas_tdr}, owing to their cost-effectiveness and excellent time resolution. Despite their long-standing use, several performance-related limitations remain poorly understood \cite{rpcML}. In particular, low-resistive bakelite RPCs operated with self-trigger electronics (XYTER) \cite{stsmuchxyter2} have been found to produce secondary (fake) hits a few nanoseconds after the primary signal \cite{Chattopadhay_2025, GANAI2023168384, MONDALMitaliRPC}. These delayed hits degrade not only the time resolution but also the spatial resolution of the detector. Increasing the ADC threshold helps suppress such fake hits to some extent but simultaneously reduces the overall detection efficiency \cite{Chattopadhay_2025}. Distinguishing these secondary hits from genuine signals is particularly difficult in self-trigger mode, where no external reference trigger is available. We have been investigating possible approaches to address this problem. Machine-learning techniques may offer a promising solution by effectively identifying and separating signal and background hits. Recent developments have demonstrated their ability to enhance timing resolution in RPCs \cite{rpcML}, support long-term detector performance monitoring \cite{cms_rpc_ml}, and efficiently discriminate signal from background in collider environments \cite{cms_sig_background}. Motivated by these advances, this work investigates machine-learning models for classifying primary and secondary hits in self-triggered RPC systems. The models are trained using features explicitly designed to be independent of any external trigger, ensuring compatibility with self-trigger operation.
The remainder of this article is organized as follows. Section~\ref{sec:ex_setup} outlines the experimental setup and data acquisition procedure. Section~\ref{sec:feature_des} presents a detailed study of intra-cluster observables for signal and background hits and describes the feature construction process. Section~\ref{sec:ai_model} introduces the machine-learning models-DNN, 1D-CNN, and BDT-along with the corresponding performance evaluation metrics. Section \ref{sec:performance} presents the performance evaluation of the models using appropriate metrics. Finally, Section~\ref{sec:summary} provides the summary and discussion.

\section{Experimental Setup}
\label{sec:ex_setup}
\begin{figure}[htb]
\centering
\includegraphics[width=0.8\linewidth]{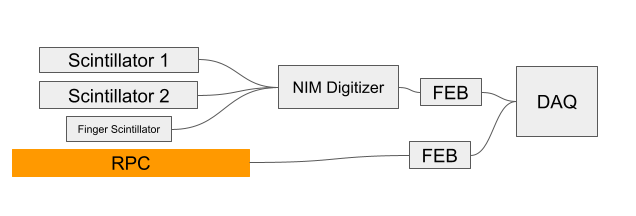}
\caption{Schematic of the laboratory experimental setup. Three scintillators and one RPC are arranged along the beam axis. Signals from the scintillators are processed by NIM modules (leading-edge discriminators and quad-coincidence logic) to generate the master trigger, while RPC signals are fed to the Front-End Boards (FEBs) and recorded by the DAQ system.}
\label{fig:expsetup}
\end{figure}

The laboratory setup used for data collection consisted of three scintillators and a single-gap RPC, as shown in Fig.~\ref{fig:expsetup}. Two large scintillators (Scintillator-1 and Scintillator-2) and a narrow finger scintillator were arranged vertically, with the RPC placed just below the finger scintillator. Signals from the three scintillators were processed through NIM electronics-using leading-edge discriminators followed by a quad-coincidence logic module-to produce a master coincidence trigger. This master trigger, along with the RPC readout signals, was sent to the Front-End Boards (FEBs), whose outputs were then digitized and recorded by the data acquisition (DAQ) system.

The readout chain consists of the Front-End Board (FEB), the FPGA-based data processing board, and the First-Level Event Selector (FLES) with its interface board (FLIB)\cite{stsmuchxyter2}.
Data were collected at a controlled temperature of \(22^{\circ}\mathrm{C}\) and a relative humidity of 40\%, with the RPC operated at a high voltage of 10.0~kV.

\subsection{Feature Construction}
\label{sec:feature_des}
\begin{figure}
    \centering
    \includegraphics[width=0.7\linewidth,
        trim=0cm 0cm 0cm 1cm, clip]{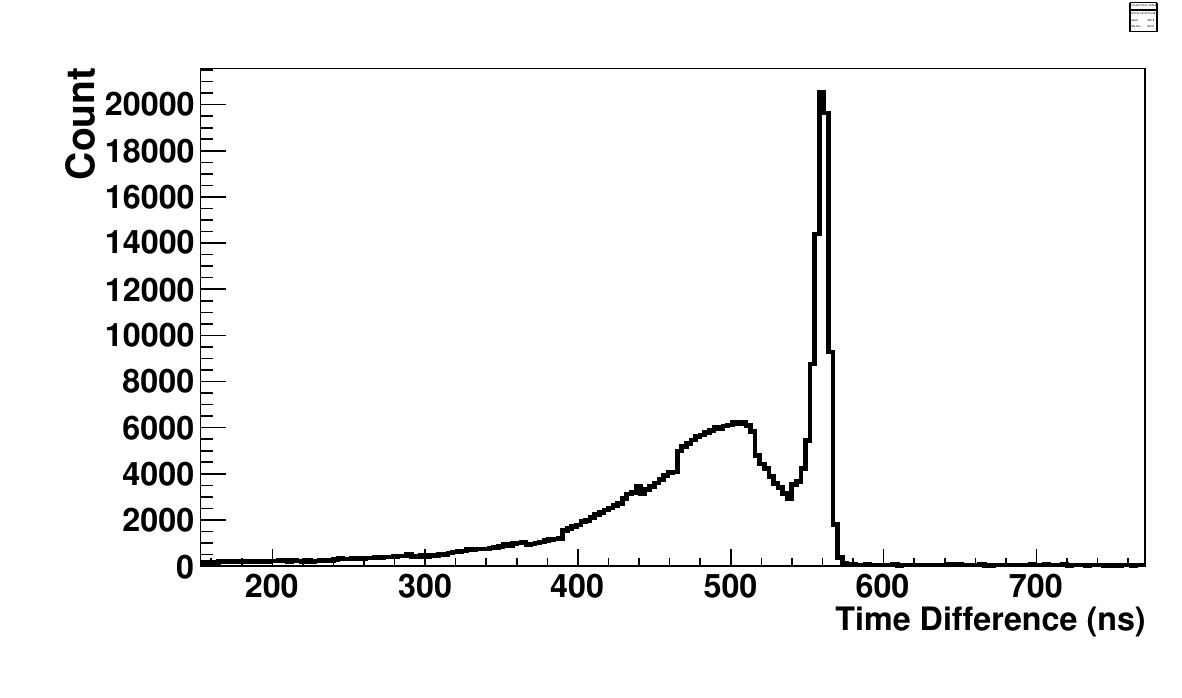}
    \caption{Time-correlation plot with respect to master trigger}
    \label{fig:time_corr_data}
\end{figure}

Figure~\ref{fig:time_corr_data} shows the time-correlation spectrum of the RPC detector with respect to the master trigger. The data reveal a distinct secondary peak in the time-difference distribution, originating from background hits that do not contribute to detector efficiency. In this work, signal and background hits were labeled based on their relative timing: hits belonging to the main peak were classified as signal, while those forming the secondary peak were labeled as background. However, in self-triggered operation-where no external timing reference exists-such separation at the individual-hit level is not feasible. To address this, all input features were designed to depend solely on detector observables. Since a single particle typically generates hits in multiple pads, our approach begins by performing spatial and temporal agglomerative clustering, and the resulting cluster-level properties are used as features. If clusters produced by genuine signals and those arising from secondary hits show systematic differences, these distinctions can be leveraged for classification. This motivated the development of a machine-learning–based method capable of identifying signal and background clusters without relying on trigger information.

\begin{figure}[htb]
    \centering
    \includegraphics[width=0.9\linewidth]{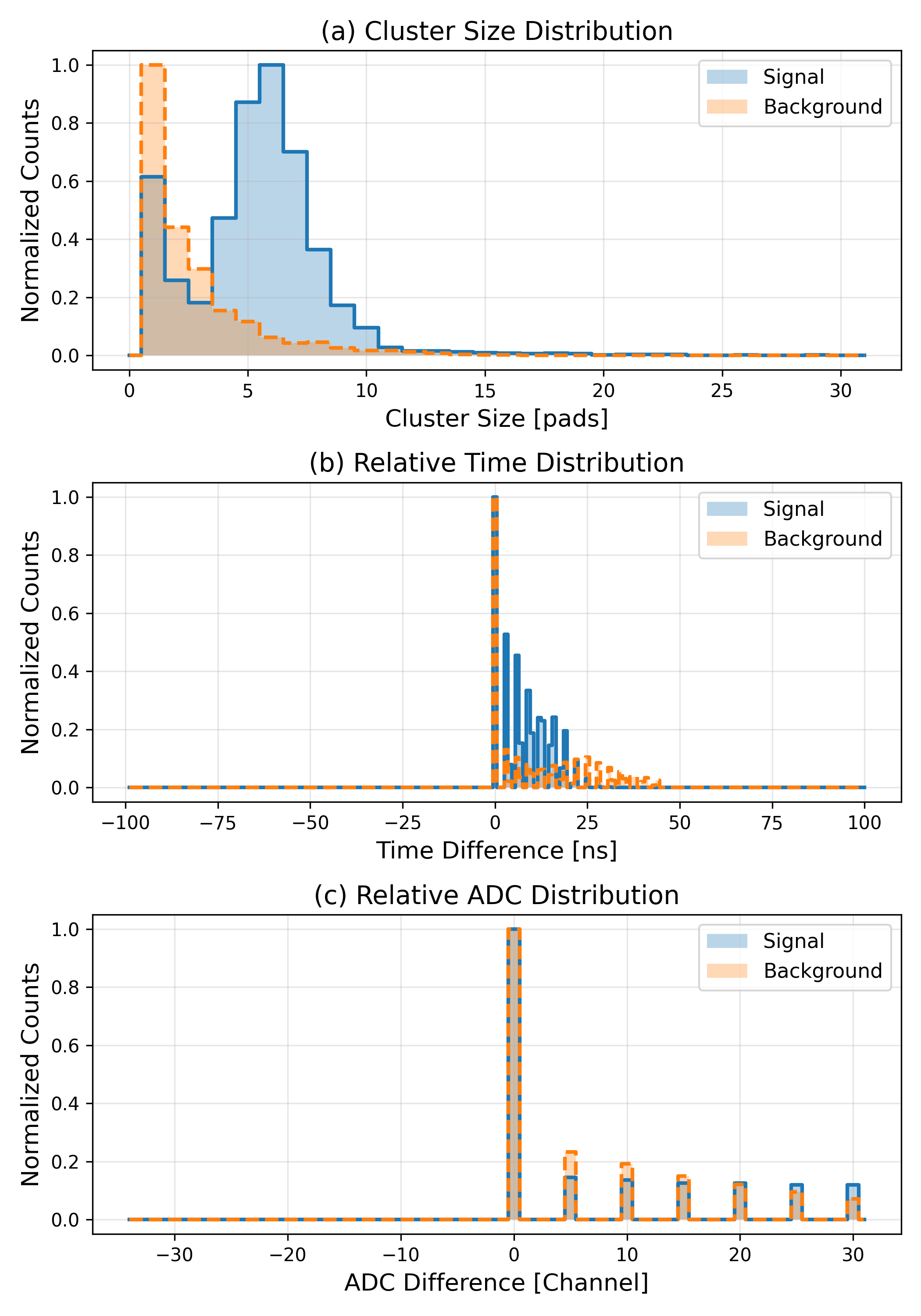}
    \caption{Comparison of signal and background clusters in terms of (a) Cluster-size (b)Intra-cluster time-difference and (c) Intra-cluster ADC-difference distributions.}
    \label{fig:comp_hists_sg_bg}
\end{figure}
To evaluate how well signal and background clusters can be distinguished, we first analyzed the distributions of various cluster-level observables. Figure~\ref{fig:comp_hists_sg_bg} compares signal and background clusters for two representative variables: time difference and ADC difference. The time difference is defined as the difference between the time of each hit and the earliest hit in the cluster, providing a measure of the cluster’s temporal spread. The ADC difference is obtained by subtracting the minimum ADC value in the cluster from the ADC values of all other hits, reflecting variations in charge deposition within the cluster.
Distinct differences between the two classes are evident, as shown in Fig.~\ref{fig:comp_hists_sg_bg}. The mean cluster size for background hits is considerably smaller than that of signal clusters. The intra-cluster time-difference distribution for signal clusters is much narrower, indicating tightly correlated hit timings. In contrast, background clusters typically exhibit broader time spreads because of uncorrelated or delayed induced signals. For the ADC-difference distribution, background clusters show a more pronounced falling tail compared to signal clusters, reflecting the presence of additional low-charge hits caused by neighboring activity. These characteristic differences form the foundation for designing discriminative features and guided the selection of input variables for the classifier.

A total of fifteen descriptive variables were selected as input features for the classification models. These features were chosen to capture both the statistical behavior and the fitted parameters of the intra-cluster time and charge (ADC) distributions, providing a comprehensive representation of each cluster. Together, they encapsulate the essential statistical and parametric properties of the time and ADC profiles associated with every cluster. The full set of input features is summarized below:

\begin{itemize}
\item Cluster size (\texttt{clustersize}): Number of adjacent fired strips forming a cluster.

\item Mean and sigma of time difference (\texttt{time\_hist\_mean}, \texttt{time\_hist\_sigma}): Mean and standard deviation of the intra-cluster time-difference distribution of the digi hits.

\item Fitted mean, sigma, and amplitude of time difference (\texttt{time\_fit\_mean}, \texttt{time\_fit\_sigma}, \texttt{time\_fit\_amp}): Mean, sigma, and peak amplitude obtained from a Gaussian fit to the intra-cluster time-difference distribution.

\item Time-difference fit quality parameters (\texttt{time\_chi2}, \texttt{time\_ndf}): Chi-square and number of degrees of freedom from the Gaussian fit, representing the quality of the time-distribution fit.

\item Mean and sigma of ADC difference (\texttt{adc\_hist\_mean}, \texttt{adc\_hist\_sigma}): Mean and standard deviation of the intra-cluster ADC-difference distribution of the digi hits.

\item Fitted mean, sigma, and amplitude of ADC difference (\texttt{adc\_fit\_mean}, \texttt{adc\_fit\_sigma}, \texttt{adc\_fit\_amp}): Mean, width, and amplitude obtained from a Gaussian fit to the intra-cluster ADC distribution, describing its shape and intensity.

\item ADC fit quality parameters (\texttt{adc\_chi2}, \texttt{adc\_ndf}): Chi-square and number of degrees of freedom from the Gaussian fit to the ADC distribution, quantifying the fit quality.
\end{itemize}
Together, these features capture the essential temporal and charge characteristics of each reconstructed cluster, reflecting both their statistical variations and the quality of the Gaussian fits. By combining histogram-based and fit-based descriptors, the feature set effectively encodes subtle shape differences between signal and background clusters. The same set of features was used for all three classifiers-Deep Neural Network (DNN), Convolutional Neural Network (CNN), and Boosted Decision Tree (BDT)-ensuring a consistent basis for comparing their performance.
\section{Machine Learning Model}
\label{sec:ai_model}
\subsection{DNN and 1D-CNN model}
A Deep Neural Network (DNN) is a feed-forward architecture consisting of multiple nonlinear layers that progressively learn higher-level representations of the input data. In contrast, a 1D Convolutional Neural Network (1D-CNN) employs convolutional filters along a single spatial dimension to capture local and hierarchical patterns in sequential inputs. Owing to weight sharing and local connectivity, CNNs are particularly effective for processing structured 1D signals \cite{lecun2015deep}. Nonlinear activation functions such as ReLU and Sigmoid enable these models to learn complex dependencies that cannot be represented by purely linear operations. The architectures of the DNN and 1D-CNN models used in this study are summarized in Tables~\ref{tab:dnn_arch} and~\ref{tab:cnn_arch}, respectively. The DNN consists of two fully connected hidden layers with ReLU activation and a dropout layer for regularization, followed by a sigmoid output layer for binary classification, with a total of 3,137 trainable parameters. In contrast, the 1D-CNN model employs convolutional layers to capture local correlations among the input features, followed by batch normalization, global max-pooling, and dropout for stabilization and regularization, and a final dense layer with sigmoid activation. The CNN architecture contains a larger number of parameters (17,857 in total), reflecting its increased representational capacity.
\label{subsec:dnn_cnn}

\begin{table}[ht]
\centering
\caption{Architecture summary of the DNN model.}
\label{tab:dnn_arch}
\begin{tabular}{lcc}
\hline
\textbf{Layer (type)} & \textbf{Output Shape} & \textbf{Number of Parameters} \\
\hline
Dense (64 neurons, ReLU) & (None, 64) & 1,024 \\
Dropout (rate = 0.2) & (None, 64) & 0 \\
Dense (32 neurons, ReLU) & (None, 32) & 2,080 \\
Dense (1 neuron, Sigmoid) & (None, 1) & 33 \\
\hline
\textbf{Total parameters} &  & \textbf{3,137} \\
\textbf{Trainable parameters} &  & 3,137 \\
\textbf{Non-trainable parameters} &  & 0 \\
\hline
\end{tabular}
\end{table}

\begin{table}[ht]
\centering
\caption{Architecture summary of the 1D CNN model.}
\label{tab:cnn_arch}
\begin{tabular}{lcc}
\hline
\textbf{Layer (type)} & \textbf{Output Shape} & \textbf{Number of Parameters} \\
\hline
Conv1D (128 filters, kernel size = 3, ReLU) & (None, 11, 128) & 768 \\
Conv1D (32 filters, kernel size = 4, ReLU) & (None, 8, 32) & 16,416 \\
Batch Normalization & (None, 8, 32) & 128 \\
Global MaxPooling1D & (None, 32) & 0 \\
Dropout (rate = 0.2) & (None, 32) & 0 \\
Dense (16 neurons, ReLU) & (None, 16) & 528 \\
Dense (1 neuron, Sigmoid) & (None, 1) & 17 \\
\hline
\textbf{Total parameters} &  & \textbf{17,857} \\
\textbf{Trainable parameters} &  & 17,793 \\
\textbf{Non-trainable parameters} &  & 64 \\
\hline
\end{tabular}
\end{table}

The classifier was trained on balanced datasets containing equal numbers of signal and background clusters to avoid bias. DNN and 1D-CNN models were implemented using the Keras API within TensorFlow \cite{tensorflow, keras}. The DNN model was trained using the Adam optimizer with a binary cross-entropy loss function. Training was performed with a batch size of 128 for a maximum of 60 epochs. Early stopping was applied by monitoring the validation AUC with a patience of 8 epochs, and the best model weights were restored. A separate validation dataset was used to track performance during training. For the 1D-CNN model, hyperparameter tuning was performed using the Keras Tuner framework to optimize the model architecture and training parameters. Each trial was trained for up to 30 epochs with a batch size of 128, and early stopping was applied with a patience of 6 epochs based on validation AUC. The best-performing hyperparameter configuration was selected for the final model. The DNN model required approximately 10 seconds for training, while the 1D-CNN model  required approximately 33 seconds.
\subsection{Boosted Decision Tree (BDT) Model}
\label{sec:bdt}
A Decision Tree is a supervised learning model that predicts outcomes by recursively splitting the feature space into regions using simple threshold-based rules, with each leaf node representing a final class or value. A Boosted Decision Tree (BDT) builds on this idea by combining many such weak decision trees into a strong ensemble using iterative boosting\cite{bdt_friedman}.
The Boosted Decision Tree model was implemented using the \texttt{XGBoost} framework\cite{xgboast}. The optimal configuration was found to consist of 200 trees with a maximum depth of 5 and a learning rate of 0.05. A row subsampling fraction of 0.7 was applied to enhance generalization and reduce overfitting. This setup provides a balanced trade-off between model complexity and stability, making it well suited for distinguishing genuine signal clusters from crosstalk-induced background in RPC detector data.
\section{Performance Evaluation}
\label{sec:performance}
\begin{figure}[htb]
    \centering
    \includegraphics[width=\linewidth]{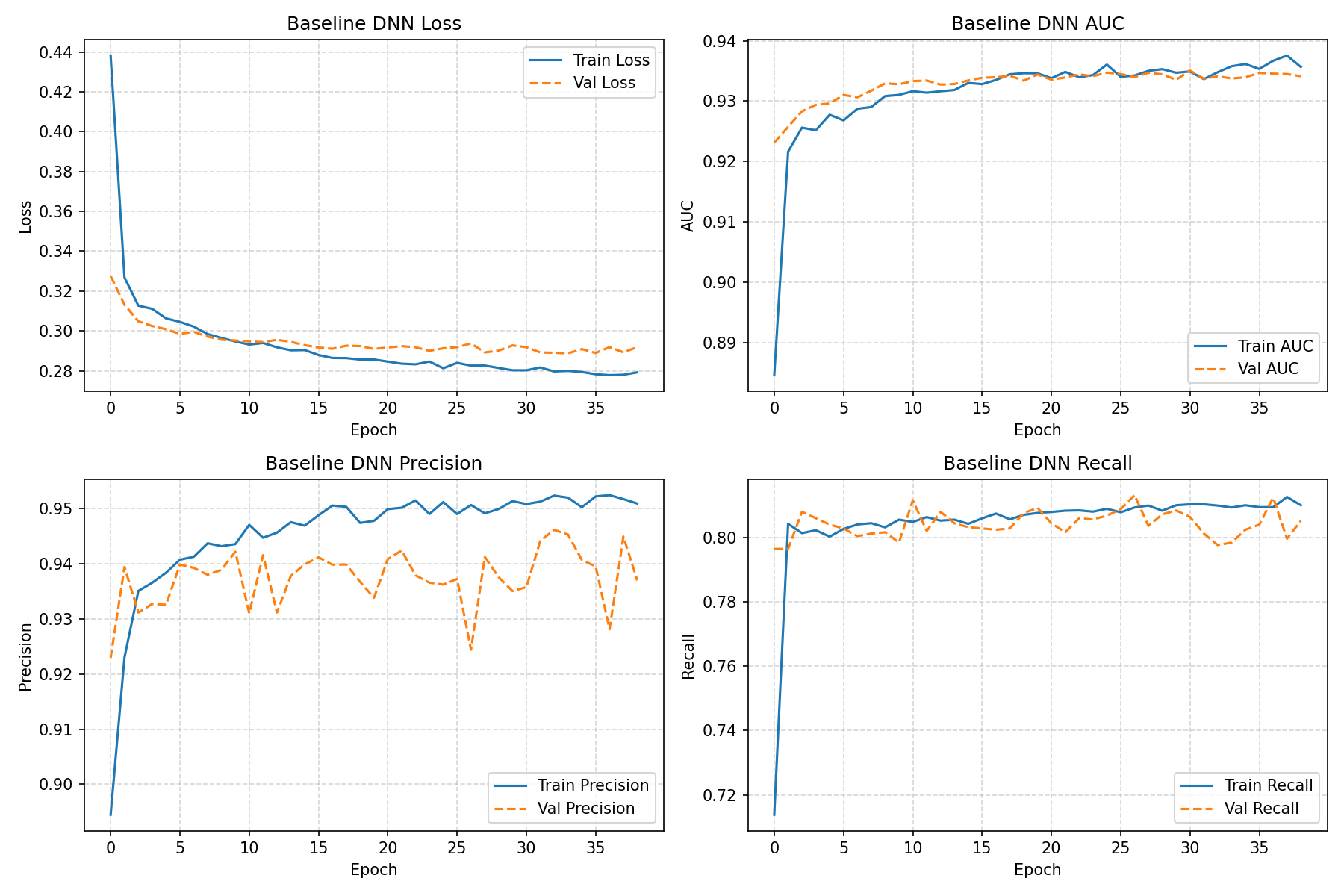}
    \caption{Training history of the baseline DNN model showing accuracy and loss evolution over epochs.}
    \label{fig:dnn_training_history}
\end{figure}
Figure \ref{fig:dnn_training_history} shows the training history of the DNN, where both training and validation loss converge smoothly without significant divergence, indicating stable training and no evident overfitting.

\begin{figure}[htb]
    \centering
    \includegraphics[width=0.6\linewidth]{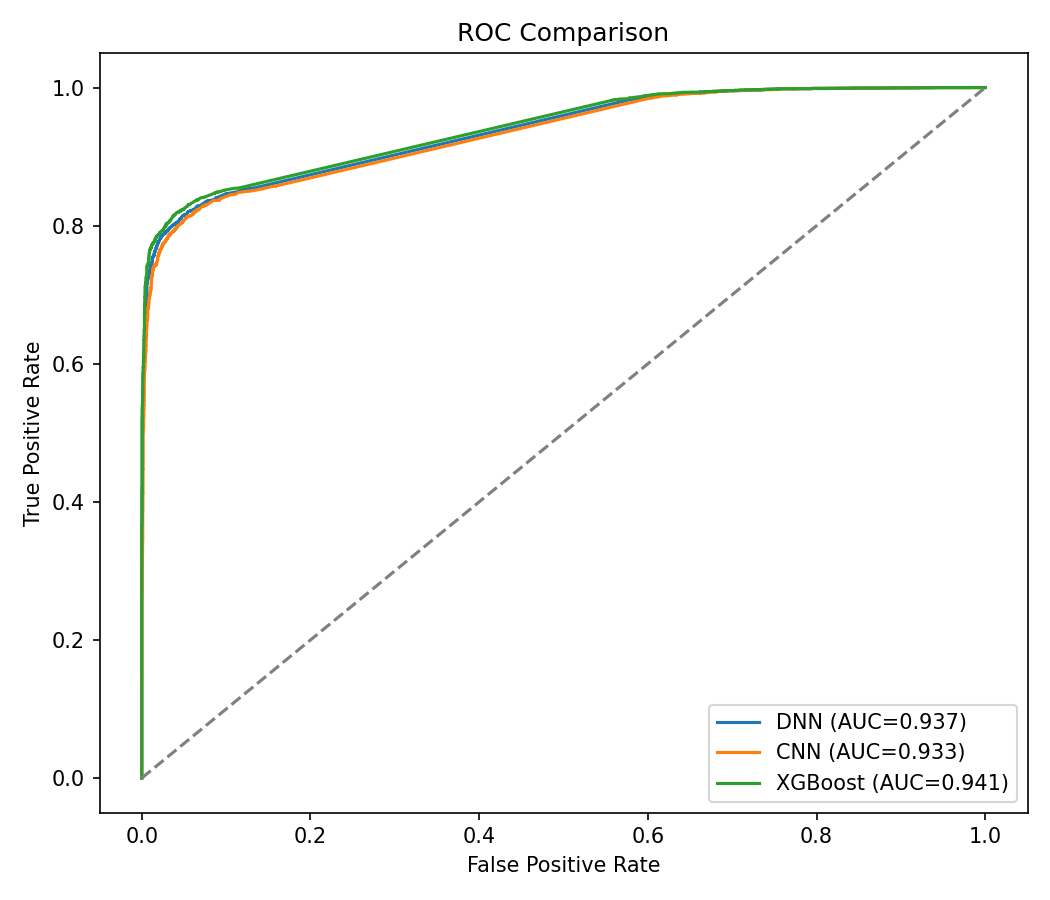}
    \caption{Comparison of ROC curves for DNN, CNN, and BDT classifiers.}
    \label{fig:rocall}
\end{figure}
Figure~\ref{fig:rocall} presents the Receiver Operating Characteristic (ROC) curves for the three classifiers-DNN, Convolutional Neural Network (CNN), and Boosted Decision Tree (XGBoost). The XGBoost achieves the highest area under the ROC curve (AUC) of 0.941, indicating strong separation power between signal and background clusters. The corresponding Precision–Recall (PRC) analysis confirms consistent precision across a wide range of recall values, which is particularly important in the background-dominated RPC environment.

\begin{table}[htbp]
\centering
\caption{Feature importances from the trained XGBoost classifier (average gain). The importance scores are normalized to sum to unity. Each feature corresponds to a specific physical or statistical property of the detected cluster.}
\label{tab:xgb_feature_importances_desc}
\begin{tabular}{lcc}
\hline
\textbf{Feature} & \textbf{Gain} & \textbf{Share (\%)} \\
\hline
clustersize & 0.4177 & 41.8 \\
time\_hist\_sigma & 0.2290 & 22.9 \\
time\_fit\_amp & 0.0936 & 9.4 \\
adc\_hist\_sigma & 0.0529 & 5.3 \\
adc\_chi2 & 0.0335 & 3.3 \\
time\_chi2 & 0.0254 & 2.5 \\
adc\_ndf & 0.0232 & 2.3 \\
adc\_hist\_mean & 0.0243 & 2.4 \\
time\_hist\_mean & 0.0171 & 1.7 \\
adc\_fit\_amp & 0.0182 & 1.8 \\
time\_fit\_mean & 0.0148 & 1.5 \\
adc\_fit\_sigma & 0.0145 & 1.5 \\
time\_fit\_sigma & 0.0144 & 1.4 \\
adc\_fit\_mean & 0.0094 & 0.9 \\
time\_ndf & 0.0119 & 1.2 \\
\hline
\textbf{Total} & \textbf{1.0000} & \textbf{100.0} \\
\hline
\end{tabular}
\end{table}

The feature-importance analysis of the XGBoost model (Table~\ref{tab:xgb_feature_importances_desc}) indicates that the \textit{cluster size} is the most influential variable, contributing over 40\% of the total gain. Temporal-shape descriptors such as the \textit{time histogram width} and \textit{time-fit amplitude} also play significant roles, together accounting for more than 30\% of the gain. Charge-distribution–related variables, including the \textit{ADC histogram} and \textit{ADC-fit parameters}, further enhance classification sensitivity. These results confirm that both timing and amplitude information are crucial for effectively distinguishing genuine signal clusters from background or crosstalk-induced hits.
\begin{figure}[htbp]
    \centering

    \includegraphics[width=0.63\textwidth]{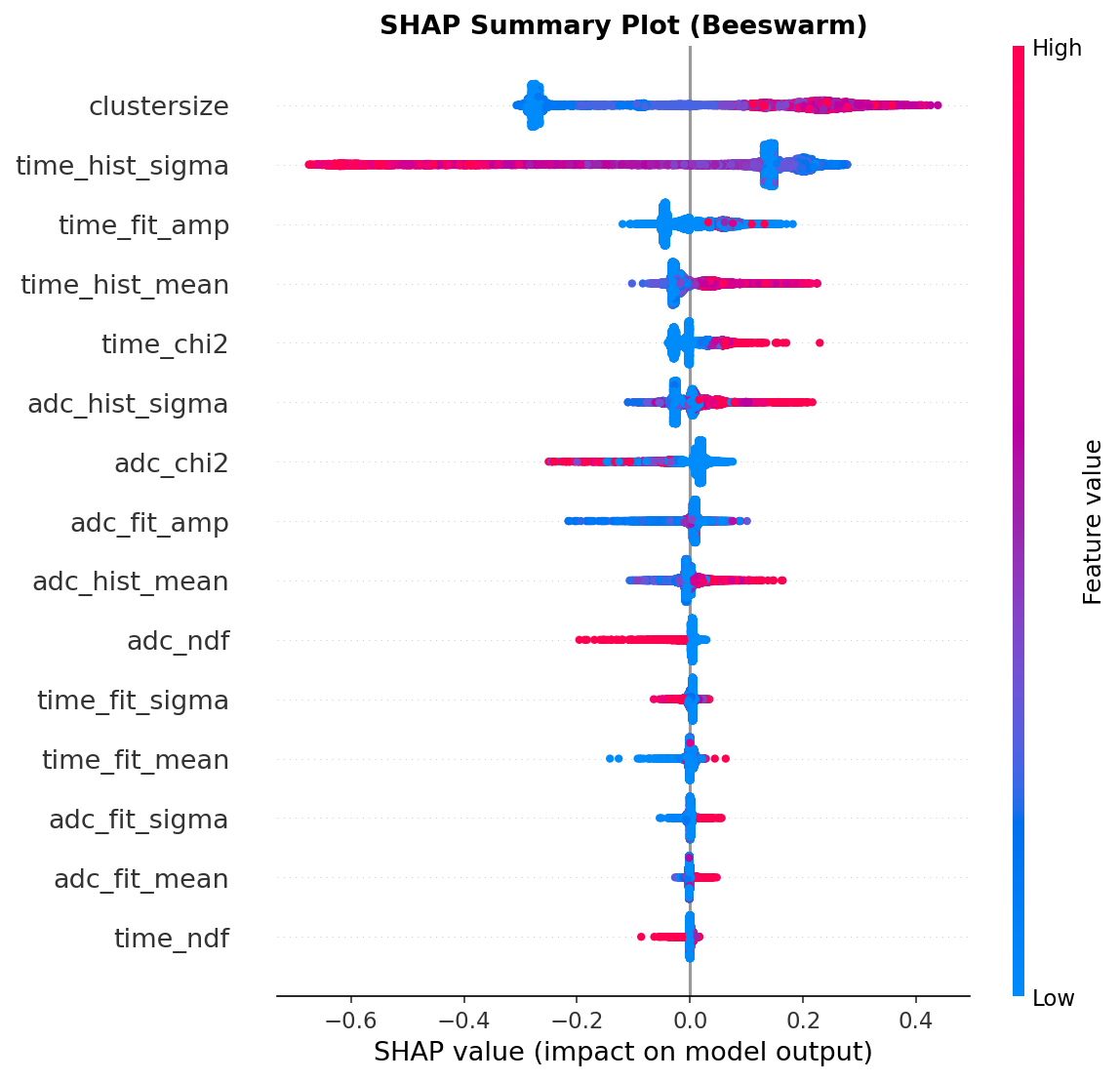}
    \caption*{(a) SHAP summary plot (beeswarm)}

    \vspace{0.5cm}

    \includegraphics[width=0.63\textwidth]{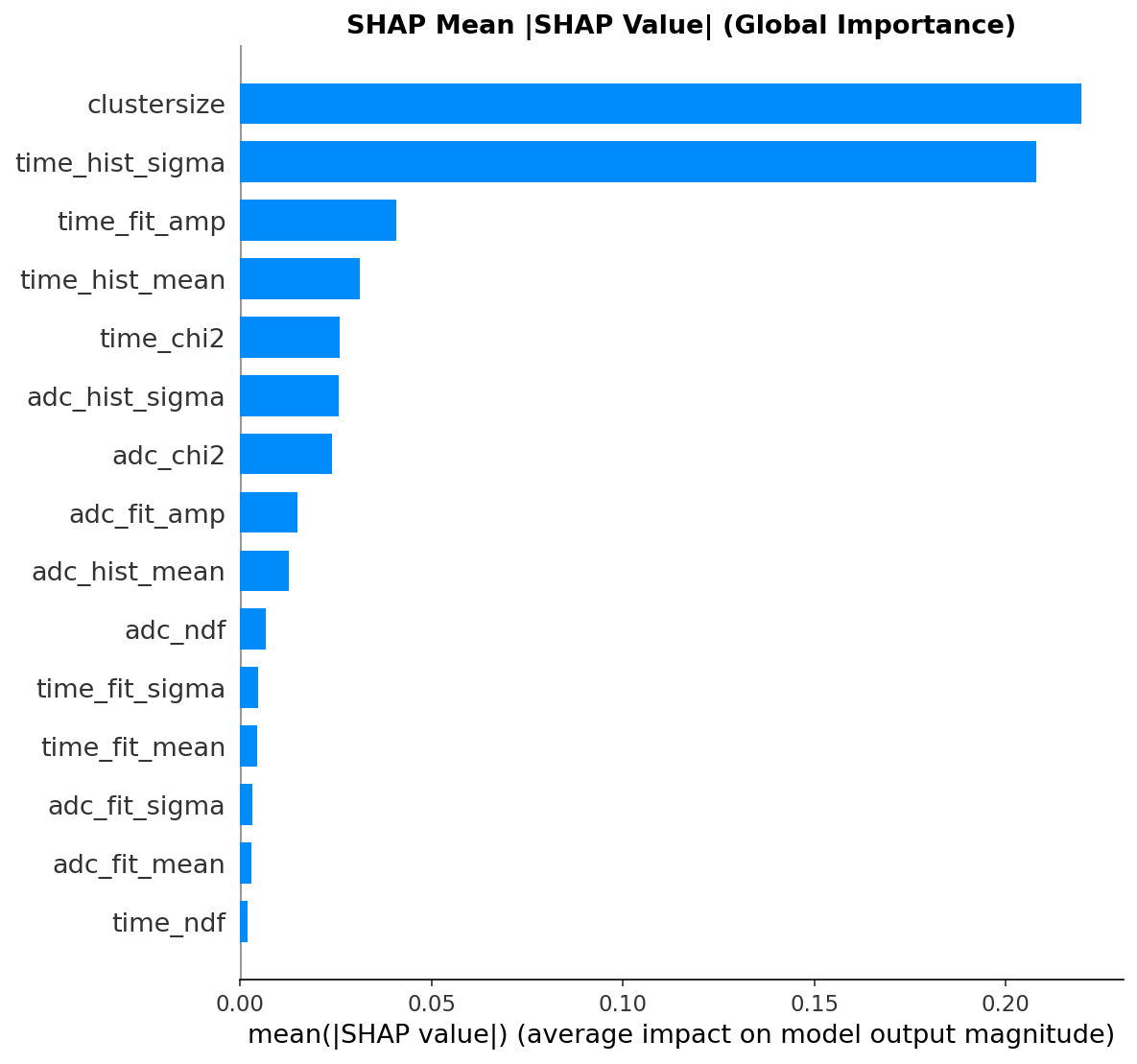}
    \caption*{(b) SHAP global importance (mean $|\mathrm{SHAP}|$)}

    \caption{SHAP-based interpretation of the XGBoost model. 
(a) The summary (beeswarm) plot shows the distribution and direction of feature contributions for individual events. The position along the x-axis indicates the SHAP value, i.e., the contribution of each feature to the model output (positive values favor signal classification, while negative values favor background). The color of the points represents the feature value, with red indicating higher values and blue indicating lower values. (b) The bar plot represents the global feature importance based on the mean absolute SHAP values, indicating the overall contribution of each feature to the model predictions.}
    
    \label{fig:shap_plots}
\end{figure}
To further investigate the contribution of input features to the XGBoost model, a SHAP (Shapley Additive Explanations) analysis was performed. The results are presented in Fig.~\ref{fig:shap_plots}. The SHAP analysis is found to be consistent with the feature importance obtained from the XGBoost gain metric, while providing additional interpretability. In particular, both approaches identify \texttt{clustersize} (cluster size) and time-related features, especially \texttt{time\_hist\_sigma} (standard deviation of the intra-cluster time-difference distribution), as the dominant discriminative variables. The SHAP summary (beeswarm) plot shows that larger cluster sizes contribute positively towards signal classification, whereas larger time spread values shift predictions towards background. This behavior is physically consistent with the expectation that true signal clusters are more compact in time. The global SHAP importance further confirms this ranking, with time-related shape parameters such as \texttt{time\_fit\_amp} (amplitude from Gaussian time fit) and \texttt{time\_hist\_mean} (mean of intra-cluster time-difference distribution) contributing significantly, while ADC-based observables such as \texttt{adc\_hist\_sigma} (standard deviation of ADC distribution) and \texttt{adc\_chi2} ($\chi^2$ of ADC fit) play a secondary role. Several features exhibit negligible contributions, indicating a degree of redundancy in the feature set. Overall, the agreement between SHAP and XGBoost importance demonstrates that the model relies on physically meaningful observables.

\begin{figure}[htb]
    \centering
    \includegraphics[width=0.95\linewidth]{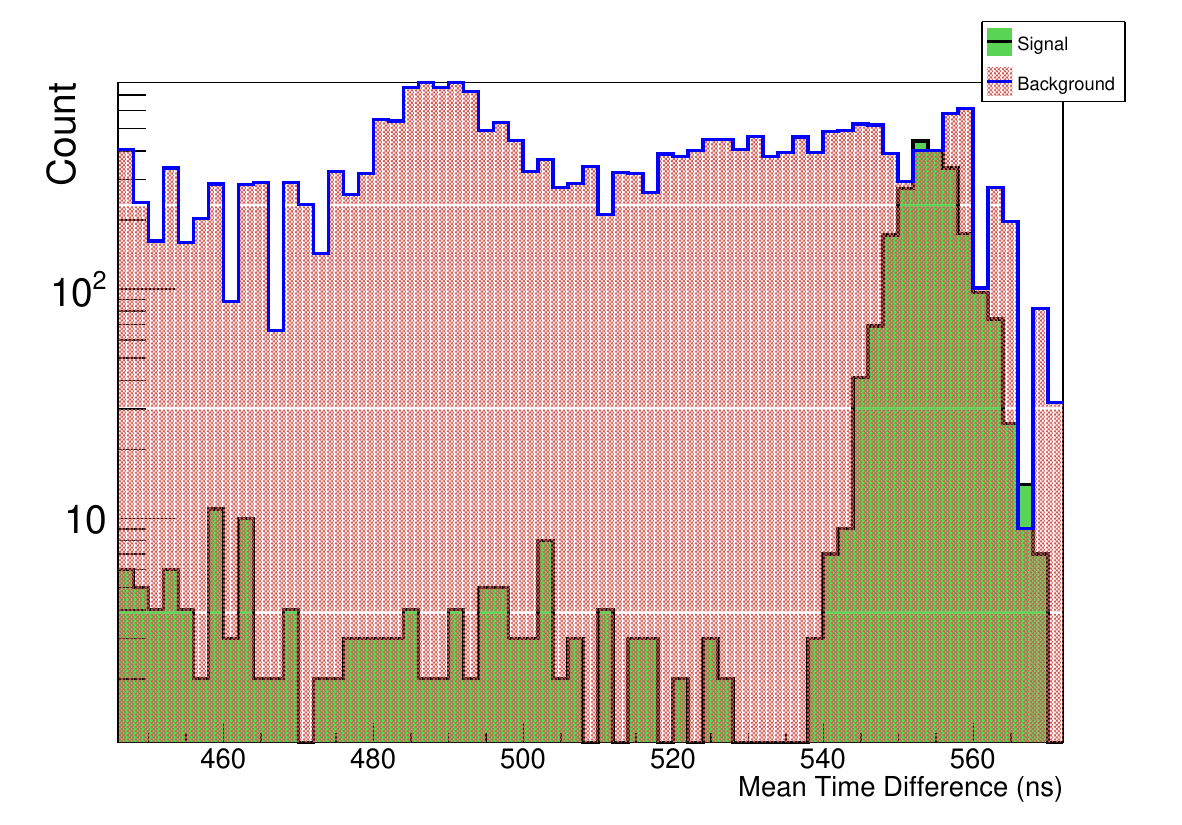}
    \caption{The X-axis shows the mean time difference between each cluster and the trigger signal. The DNN output on real data exhibits a clear separation between signal clusters forming the main peak and background clusters forming the secondary peak.}
    \label{fig:sig_bg_separation_real}
\end{figure}

Figure~\ref{fig:sig_bg_separation_real} illustrates the performance of the DNN classifier on the test dataset. 
The model correctly assigns the majority of main-peak clusters to the signal class and secondary-peak clusters to the background class. 
This demonstrates that the network effectively suppresses crosstalk-induced background events that cannot be removed through simple ADC threshold cuts.

\begin{table}[ht]
\centering
\caption{Performance comparison of the DNN, CNN, and XGBoost classifiers on the independent test dataset.}
\label{tab:model_comparison}
\begin{tabular}{lccccc}
\hline
\textbf{Model} & \textbf{AUC} & \textbf{Accuracy} & \textbf{Precision} & \textbf{Recall} & \textbf{F1 Score} \\
\hline
DNN & 0.937 & 0.882 & 0.946 & 0.811 & 0.873 \\
CNN & 0.933 & 0.879 & 0.942 & 0.807 & 0.870 \\
XGBoost & \textbf{0.941} & \textbf{0.888} & \textbf{0.954} & \textbf{0.816} & \textbf{0.880} \\
\hline
\end{tabular}
\end{table}

Table~\ref{tab:model_comparison} summarizes the quantitative performance metrics of the DNN, CNN, and XGBoost classifiers on the independent test dataset. All three models exhibit strong discrimination capability between signal and background clusters, with AUC values exceeding 0.93. Using bootstrap resampling, the AUC scores are 0.9367 ± 0.0021 (DNN), 0.9333 ± 0.0022 (CNN), and 0.9405 ± 0.0021 (XGBoost). These results indicate comparable performance within statistical uncertainties, with XGBoost achieving a marginally higher AUC and F1 score, suggesting slightly better generalization.
\begin{figure}[htbp]
    \centering
    
    \begin{minipage}[t]{0.9\textwidth}
        \centering
        \includegraphics[width=\textwidth]{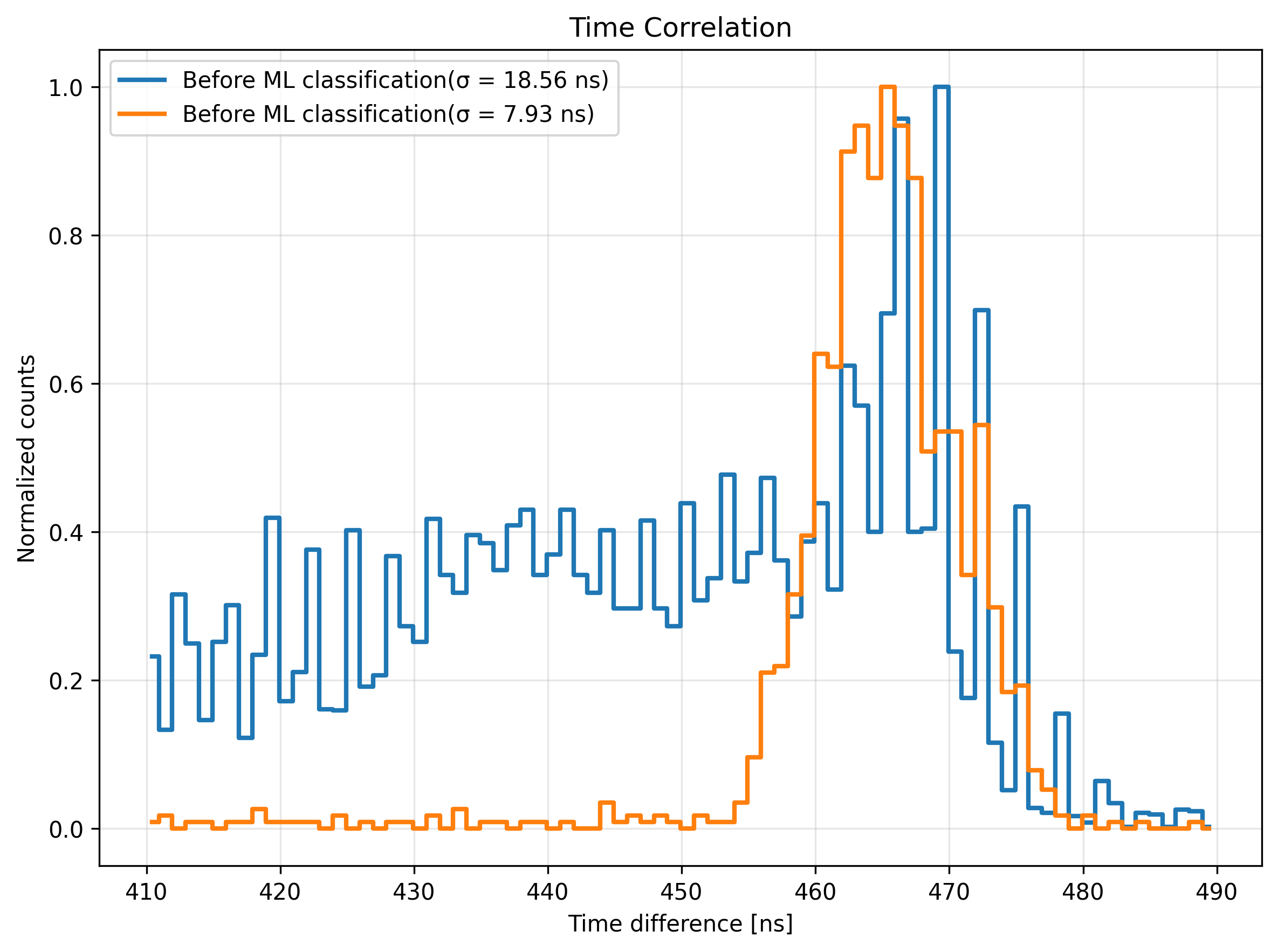}
        \caption*{(a) Time correlation distribution}
    \end{minipage}
    \vfill
    \vspace{.5cm}
    \begin{minipage}[t]{0.9\textwidth}
        \centering
        \includegraphics[width=\textwidth]{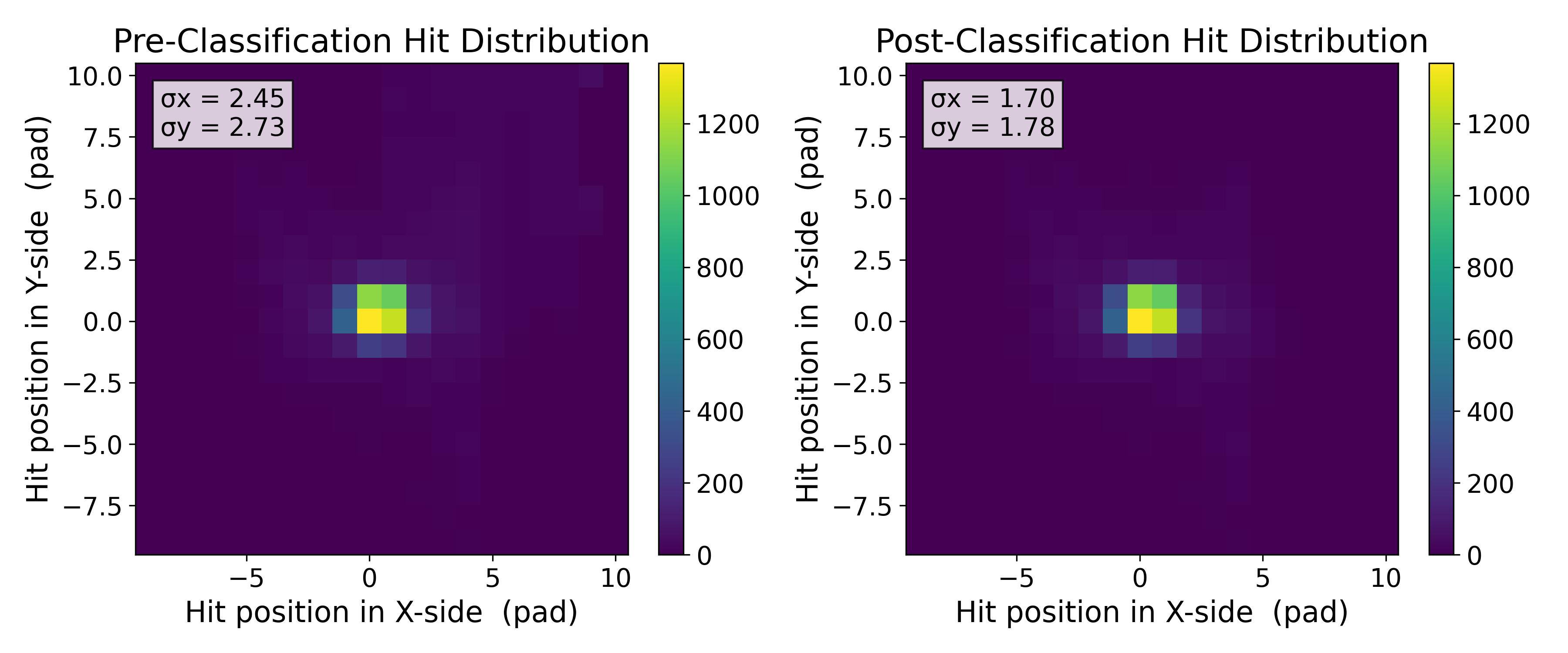}
        \caption*{(b) Spatial hit distribution}
    \end{minipage}
    
   \caption{
Impact of the machine-learning-based selection on system performance. 
(a) Time-correlation distributions before and after classification. 
(b) Two-dimensional hit position distributions before and after classification, with hits centered at (0,0).
}
    
    \label{fig:resolution_improvement}
\end{figure}
To evaluate the impact of the proposed machine-learning-based selection on detector performance, we study both the time and spatial distributions of reconstructed hits(cluster), as shown in Fig.~\ref{fig:resolution_improvement}. The total time resolution of the system can be expressed as
\[
\sigma_{\text{total}}^{2} = \sigma_{\text{RPC}}^{2} + \sigma_{\text{Sc}}^{2} + \sigma_{\text{TDC}}^{2},
\]
where $\sigma_{\text{RPC}}$, $\sigma_{\text{Sc}}$, and $\sigma_{\text{TDC}}$ represent the intrinsic time resolutions of the RPC, scintillator, and readout electronics, respectively. In this study, we compare the total time resolution ($\sigma_{\text{total}}$) before and after applying the machine-learning-based selection. The time-correlation distributions before and after classification are presented in Fig.~\ref{fig:resolution_improvement}(a). Before applying the model, the distribution is broader and exhibits a significant contribution from delayed hits, resulting in a larger spread. After classification, the distribution becomes noticeably narrower and more sharply peaked. The standard deviation ($\sigma$) decreases from 18.56~ns (before classification) to 7.93~ns (after classification). This substantial reduction indicates that delayed and background hits are effectively suppressed by the model, leading to an improvement in the effective time resolution. 

In the case of spatial resolution, direct measurement typically requires a dedicated experimental setup and reconstruction techniques. In the absence of such measurements, we infer the improvement in spatial resolution from the spatial distribution of reconstructed hits. The two-dimensional hit distributions are projected along the X and Y directions, with the cluster centers aligned at $(0,0)$, as shown in Fig.~\ref{fig:resolution_improvement}(b). The width of these projected distributions, characterized by their standard deviation ($\sigma$), provides a measure of the spatial spread. After applying the model, a clear reduction in the spread is observed. The standard deviations decrease from $\sigma_x = 2.45$ to 1.70 and from $\sigma_y = 2.73$ to 1.78, indicating improved localization of hits and, consequently, an enhancement in the effective spatial resolution. Overall, these results demonstrate that the proposed approach not only improves classification performance but also leads to measurable improvements in key detector observables.

We performed a dedicated benchmark test to evaluate the computational performance of the full analysis pipeline, including clustering, feature extraction, and XGBoost inference. The benchmark was carried out on a system equipped with an AMD EPYC 7713 64-Core processor, and the current implementation was executed in single-threaded mode in order to evaluate the intrinsic performance of the algorithm. In the benchmark test, a total of 100,000 events were processed. For our detector, a muon passing through the detector produces on average about 15 hits (digits), including both signal and background contributions. Therefore, the benchmark corresponds to processing approximately $1.5\times10^{6}$ hits in total. The full dataset was processed in $4.46~\mathrm{s}$, corresponding to an average processing time of $44.6~\mu\mathrm{s}$ per event and $2.98~\mu\mathrm{s}$ per hit, with a memory usage of approximately $330~\mathrm{MB}$. A detailed breakdown of the processing time shows that the clustering stage requires about $19.4~\mu\mathrm{s}$ per event, feature extraction about $2.4~\mu\mathrm{s}$ per event, and the XGBoost inference step about $14.6~\mu\mathrm{s}$ per event. This demonstrates that the machine-learning inference introduces only a modest computational overhead and that the overall method remains computationally lightweight.

\section{Summary and Discussion}
\label{sec:summary}
A machine-learning-based approach has been developed to classify signal and background clusters in prototype Resistive Plate Chamber (RPC) detectors. The method uses fifteen cluster-level descriptors that characterize both statistical and fit-based properties of the time and charge (ADC) distributions. These features effectively capture the subtle distinctions between genuine signal clusters and spurious background clusters arising from crosstalk or overlapping hits. Three classifiers—DNN, CNN, and XGBoost—were trained and evaluated. The corresponding area-under-ROC (AUC) scores are 0.9367 $\pm$ 0.0021, 0.9333 $\pm$ 0.0022, and 0.9405 $\pm$ 0.0021, respectively. The comparable performance of the BDT and neural-network-based models suggests that both approaches capture similar levels of complexity present in the data. This may indicate that the classification performance is limited by the degree of overlap between signal and background distributions in the current feature space, rather than solely by the learning capacity of the individual models, although it does not exclude the possibility that additional features could further improve discrimination. XGBoost achieves the highest AUC among the three models; however, the improvement over the neural-network approaches is small and comparable to the estimated statistical uncertainties, indicating that all models exhibit similar overall performance within the current statistical precision. The slightly better performance of XGBoost is consistent with the ability of gradient-boosted decision trees to effectively model non-linear correlations in structured feature spaces. Because the trained XGBoost model depends solely on compact cluster-level parameters, it can be readily integrated into real-time reconstruction pipelines. The method is computationally efficient and suitable for low-latency event processing, making it well suited for near real-time fake-hit identification and integration into fast data-processing or online monitoring pipelines in high-rate experiments. 

Overall, the study demonstrates that machine-learning classifiers provide improved signal–background discrimination compared to traditional amplitude-threshold-based methods. In particular, the XGBoost and DNN models offer robust and scalable solutions for improving hit selection and mitigating crosstalk-related background in self-triggering RPC detectors.

\section{Acknowledgements}
We gratefully acknowledge the Department of Atomic Energy, Govt of India, for funding the
project. Special appreciation is also due to Ganesh Das, Tushar Kanti Das for their unwavering support and assistance during the testing phase.

 \bibliographystyle{elsarticle-num} 
 \bibliography{cas-refs}

\end{document}